\documentclass[fleqn,usenatbib]{mnras}

\usepackage{newtxtext,newtxmath}
\usepackage[T1]{fontenc}

\DeclareRobustCommand{\VAN}[3]{#2}
\let\VANthebibliography\thebibliography
\def\thebibliography{\DeclareRobustCommand{\VAN}[3]{##3}\VANthebibliography}

\usepackage{graphicx}	% Including figure files
\usepackage{amsmath}	% Advanced maths commands
\usepackage{soul}       % to highlight things (reminder for edit)
\usepackage{cleveref}   % for referencing multiple equations at once

\crefname{equation}{equation}{equations}

\title[Crystallized white dwarfs]{Oscillations and tidal deformations of crystallized white dwarfs}

\author[Y.-T. Tang and L.-M. Lin]{
Yat-To Tang,$^{1}$\thanks{E-mail: yttang@phy.cuhk.edu.hk}
Lap-Ming Lin$^{1}$
\\
$^{1}$Department of Physics, The Chinese University of Hong Kong, Hong Kong, China
}

\date{Accepted XXX. Received YYY; in original form ZZZ}

\pubyear{2015}

\begin{document}
\label{firstpage}
\pagerange{\pageref{firstpage}--\pageref{lastpage}}
\maketitle

\begin{abstract}
Long predicted more than fifty years ago, strong evidence for the existence of crystalline cores 
inside white dwarfs has recently been obtained by the Gaia space telescope. It is thus important to investigate how 
a crystalline core may affect the properties and dynamics of white dwarfs. 
In this paper, we first study the dependence of the frequencies of the fundamental (f), interfacial (i), and shear (s) oscillation modes on the size of the crystalline core. 
We find that the frequencies of the i- and s-modes depend sensitively on the size of the core, while the frequency of the f-mode is affected only slightly by at most a few percent for our chosen white dwarf models. 
We next consider the tidal deformability of crystallized white dwarfs and find that the effect of crystallization
becomes significant only when the radius of the core is larger than about 70\% of the stellar radius. 
The tidal deformability can change by a few to about 10 percent when a white dwarf becomes fully crystallized. 
We also show that there exist approximate equation-of-state insensitive relations connecting the mass, moment of inertia, tidal deformability, and f-mode frequency for pure fluid white dwarfs. Depending on the 
stellar mass and composition, however, these relations can be affected by a few percent when the white dwarf is crystallized. 
These changes could leave an imprint on the gravitational waves emitted from the late inspiral or merger of white dwarf binaries, which may be detectable by future space-borne gravitational wave detectors.
%Our work suggests that the properties of crystallization inside white dwarfs may be inferred by the gravitational wave signals. 
\end{abstract}


%%%%%%%%%%%%%%%%%%%%%%%%%%%%%%%%%%%%%%%%%%%%%%%%%%

%%%%%%%%%%%%%%%%% BODY OF PAPER %%%%%%%%%%%%%%%%%%

\section{Introduction}

White dwarfs (WDs) are among the most common stellar objects in the universe. It is estimated that our galaxy already hosts $\sim 10^8$ double white dwarf (DWD) binaries \citep{Nelemans2001}. Many of these binary systems are compact enough to be able to merge within the age of the universe. Furthermore, it is expected that gravitational wave (GW) signals emitted from DWD systems are detectable by future space-borne gravitational wave detectors, such as the Laser Interferometer Space Antenna \citep[LISA,][]{LISA2022}, the Deci-hertz Interferometer Gravitational Wave Observatory \citep[DECIGO,][]{DECIGO2021}, TianQin and Taiji \citep{
Ruan:2020v4,Gong2021}. Population synthesis models suggest that more than $10^4$ DWDs could be resolved by LISA \citep{Korol2017,Lamberts2019,Li2020,Breivik2020}.

During the inspiral of a binary system, the oscillation modes of the stars may be excited, affecting the evolution of the system by transferring the orbital energy to the oscillation modes \citep{Fuller2011,Rathore2005} or even contributing directly to the GW emission.
Although the effect on GW signal is small for DWD binaries, it is likely measurable by LISA for systems with high signal-to-noise ratio \citep{Taylor2020}. Furthermore, it is expected that DECIGO will detect 6600 DWD mergers per year \citep{Kinugawa2022}. The influence of mode excitation would be relevant for  
the GW signals emitted during the late inspiral and merger of these binaries. 
The oscillation modes of WDs have been widely investigated, focusing mostly on their gravity modes \citep[see][and references therein]{Corsico2020}. The tidal excitation of oscillation modes for binary WD systems has also been studied by \citet{Fuller2011,Fuller2012}.

Meanwhile, the fundamental (f) mode frequency and tidal deformability of neutron stars are found to be connected by the f-Love relation \citep{Chan2014}, which is approximately equation-of-state (EOS) insensitive. 
The moment of inertia, tidal deformability, and rotational induced quadrupole moment of neutron stars are also connected by the I-Love-Q universal relations \citep{ILoveQ1,ILoveQ2}. These relations can be useful tools to infer the physical quantities of neutron stars. Furthermore, the I-Love-Q and 
f-Love relations have been used with the multiple Love relations \citep{Yagi:2014} 
to reduce the intrinsic matter parameters in theoretical gravitational waveform models for binary neutron star inspirals that take into account the quadrupolar and octopolar adiabatic and dynamical tidal effects \citep{Lackey:2019,Schmidt:2019,Andersson:2019,Barkett:2020}.  
The I-Love-Q relations have recently been extended to WDs \citep{Boshkayev2017}, though the relations are altered under differential rotation \citep{Taylor2020} and could be broken for hot WDs \citep{Boshkayev2018}. 
On the other hand, the f-Love relation has not been studied for WDs as far as we are aware. 
The universal relations for WDs are also relevant to the GW measurements of DWD binaries. 
By using the I-Love relation with another relation that connect the mass and moment of inertia of WDs, it has been proposed that the individual masses of a WD binary can be measured by considering the finite-size effects of the GW signal emitted by the system \citep{Wolz2021}. 

Nevertheless, observable DWD binaries may have evolved long enough to be fully or partially crystallized. More than fifty years ago, \citet{vanHorn1968} has predicted the occurrence of crystallization during WD cooling, suggesting that it may slow down the cooling process by the release of latent heat.
Indirect evidence for the slowdown due to crystallization was observed for WDs in globular clusters 
\citep{Winget2009,Berro:2010,Campos2016}.
However, it is only recently that strong observational evidence for the existence of an isolated crystallization sequence of WDs in the Hertzsprung-Russell diagram has been obtained by the Gaia space telescope \citep{Tremblay2019}. 

As the core crystallization of WDs is now well established, it is important to understand how it may affect the properties and dynamics of WDs. 
The oscillation modes of crystallized WDs have been studied mainly for the gravity mode \citep{Montgomery1999,Corsico2005,DeGeronimo2019,Althaus2021}. Recently, \citet{Perot2022} have also shown that the tidal Love number of a crystallized WD can differ by a few percent comparing to its pure fluid counterpart. 
In this work, we shall first investigate how the core crystallization affects 
the f-mode, the interfacial (i) mode associated to the core-envelope interface, the
shear (s) mode which is mainly confined in the solid core, and the tidal Love number for nonrotating WDs. We then establish the I-Love and f-Love relations for pure fluid WD models and examine how crystallization affects the validity of these relations.

The paper is organized as follows. In Section~\ref{sec:model}, we introduce the WD model used in this paper as well as the treatment of crystallization. In Section~\ref{sec:perturbation_eqs}, we review the perturbation equations for determining the oscillation modes and tidal Love numbers. The numerical schemes for solving the perturbation equations are outlined in Section~\ref{sec:numerical_scheme}. The numerical results are presented in Section~\ref{sec:results}. Finally, we summarize and conclude our findings in Section~\ref{sec:conclusions}.
Throughout the paper, we assume geometric units with $G=c=1$ unless otherwise specified.

\section{Stellar Model}
\label{sec:model}

We consider crystallized WD models consisting of a solid (inner) core and a surrounding fluid envelope. 
For the majority of the results presented in this study, the EOS given by \citet{Chamel2016} is adopted.
The mass density $\rho$ and pressure $P$ are given by 
\begin{align}
    \rho ={}& \rho_N + \frac{m_e}{8\pi^2\lambda_e^3}\left[x_r(1+2x_r^2)\sqrt{1+x_r^2} - \ln{\left(x_r+\sqrt{1+x_r^2}\right)}\right] \notag \\
    & \times \left(1+\frac{\alpha}{2\pi}\right) - n_em_e + C_M\left(\frac{4\pi}{3}\right)^{1/3}\frac{e^2}{c^2}n_e^{4/3}Z_\text{eff}^{2/3} , \label{eq:eos1} \\
    P ={}& \frac{m_ec^2}{8\pi^2\lambda_e^3}\left[x_r\left(\frac{2}{3}x_r^2-1\right)\sqrt{1+x_r^2} + \ln{\left(x_r+\sqrt{1+x_r^2}\right)}\right] \notag \\
    & \times \left(1+\frac{\alpha}{2\pi}\right) + \frac{C_M}{3}\left(\frac{4\pi}{3}\right)^{1/3}e^2n_e^{4/3}Z_\text{eff}^{2/3} \,, \label{eq:eos2}
\end{align}
where $\rho_N$ is the mass density of the nuclei, $m_e$ is the mass of electron, $n_e$ is the number density of electron, $C_M=-0.895929255682$ is the Madelung constant \citep{Baiko2001}, $\alpha$ is the fine structure constant, $e$ is the elementary charge, $c$ is the speed of light, $x_r=\lambda_ek_e$ is a dimensionless relativity parameter with $\lambda_e$ being the electron Compton wavelength and $k_e$ being the electron Fermi wave number; $Z_\text{eff}=Z\sigma(Z)^{3/2}$ is the effective nuclear charge for nuclei with 
atomic number $Z$, where 
\begin{equation}
    \sigma(Z) = 1+\alpha \frac{12^{4/3}}{35\pi^{1/3}} \left(1-1.1866Z^{-0.267} 
     +0.27Z^{-1} \right) Z^{2/3} .
\end{equation}

The EOS corresponds to a zero-temperature Coulomb plasma with atomic nuclei of single species. The hydrogen, helium layers and the atmosphere are ignored as they contribute not more than $\sim 1$\% of the total mass \citep{Saumon2022}. The type of nuclei in the core of WD may be carbon, oxygen, helium \citep{Nelemans1998,Liebert2004,Benvenuto2005}, neon, magnesium \citep{Nomoto1984}, or even iron \citep{Isern1991,Jordan2012}. The EOS is effectively
the Chandrasekhar model \citep{Chandrasekhar1931,Chandrasekhar1935,Chandrasekhar1939} with extra effects implemented, including the lattice, electron exchange, and screening corrections. 
Besides considering this EOS model with different atomic nuclei, we shall also consider the BPS EOS
\citep{BPS1971} and the original Chandrasekhar EOS \citep{Chandrasekhar1931,Chandrasekhar1935,Chandrasekhar1939} when we examine the sensitivity of universal relations with respect to different EOS models. 
Although the BPS model is not a realistic model for white dwarfs, the inclusion of
it can test for the robustness of universal relations.

Assuming the solid core to be isotropic, the elastic properties of the core are described by the effective shear modulus \citep{Chugunov2021}
\begin{equation}
    \mu = 0.119457234091 \left( \frac{4\pi}{3} \right)^{1/3}  Z^{2/3} e^2 n_e^{4/3}   . 
    \label{eq:shear_modulus}
\end{equation}
Note that this expression only gives an upper limit  of $\mu$ as it is obtained from a Voigt average \citep{Kobyakov2015}.

\begin{figure}
    \centering
    \includegraphics[width=\columnwidth]{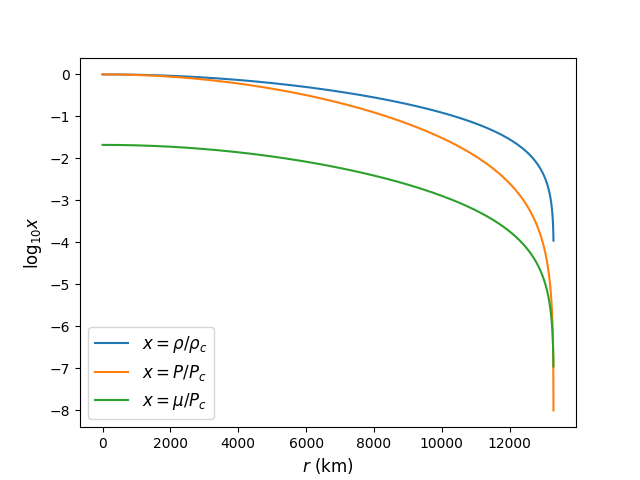}
    \caption{The profiles of density $\rho$, pressure $P$, and shear modulus $\mu$ for a $0.20M_\odot$ WD model made of $^{16}$O. The density and pressure are normalized by their central values, while the shear modulus is normalized by the central pressure. }
    \label{fig:profile}
\end{figure}

The nonrotating background stellar model is obtained by integrating the hydrostatic equilibrium equations
in Newtonian gravity
\begin{align}
    \frac{dm}{dr} &= 4\pi\rho r^2 , \label{eq:hydrostatic1} \\
    \frac{dP}{dr} &= -\rho g \,, \label{eq:hydrostatic2}
\end{align}
where $m(r)$ is the mass enclosed inside the radial coordinate $r$ and $g=m/r^2$ is the local gravitational
acceleration. The system is closed by a given EOS model $P(\rho)$. 
As an illustration of a typical WD model in our study, we plot the profiles of $\rho$, $P$, and $\mu$ for 
a $0.20M_\odot$ WD made of $^{16}$O in Figure~\ref{fig:profile}.

\section{Perturbation Equations}
\label{sec:perturbation_eqs}

In this section, we review the perturbation equations which are solved to obtain the oscillation modes and tidal Love number of our crystallized WD models.
The motion of a mass element is governed by the mass conservation equation, momentum equation, and Poisson 
equation:
\begin{align}
    &\partial_t\rho + \partial_i(\rho v_i) = 0 , \label{eq:euler1} \\
    &\partial_tv_i + v_j\partial_jv_i = \frac{1}{\rho}\partial_j\sigma_{ij} - \partial_i\Phi , \label{eq:euler2} \\
    &\partial_i\partial_i\Phi = 4\pi\rho \,, \label{eq:euler3}
\end{align}
where the Latin indices $(i,j)$ refer to the spatial coordinates, $\rho$ is the mass density, $v_i$ is the velocity of a mass element, $\sigma_{ij}$ is the stress tensor, and $\Phi$ is the gravitational potential. 
To study the oscillation modes of a given background stellar model, the above equations are
perturbed by assuming $q = q_0 + \delta q$ and keeping only terms of first order in $\delta q$, 
where $q$ denotes any physical variable such as $\rho$ and $q_0$ is its value evaluated on the unperturbed background model. Hereafter $\delta q$ is used to represent the Eulerian perturbation of $q$ and $\Delta q$ is the Lagrangian perturbation.

We assume the unperturbed background star to be in hydrostatic equilibrium without rotation, and hence $\vec{v}_0=0$ and $\delta v_i = \partial_t u_i$, where $u_i$ is the Lagrangian displacement. 
The equilibrium background is also assumed to be an inviscid fluid under zero strain, 
and the stress is only given by the isotropic pressure, i.e. $\sigma_{0,ij} = -P_0 \delta_{ij}$, where $\delta_{ij}$ is the Kronecker delta. For the treatment of elasticity, any nonzero strain is produced by the change of stress at the perturbative level. Assuming adiabatic motion for the isotropic elastic solid core, the Lagrangian perturbation of the stress tensor is given by
\begin{equation}
    \Delta\sigma_{ij} = (\Gamma_1 P_0 u_{kk})\delta_{ij} + 2\mu(u_{ij}-\frac{1}{3}u_{kk}\delta_{ij}) \,, \label{eq:stress_tensor}
\end{equation}
where $\Gamma_1$ = $d\ln{P_0}/d\ln{\rho_0}$, $\mu$ is the shear modulus, and $u_{ij}$ is the strain tensor, defined by $u_{ij} = \frac{1}{2}(\partial_j u_i + \partial_i u_j)$. As a result, the set of perturbed dynamical equations can be written as 
\begin{align}
    &\delta\rho + \partial_i(\rho u_i) = 0 , \label{eq:lin_euler1} \\
    &\rho\partial_t^2 u_i = \partial_j\delta\sigma_{ij} - \rho\partial_i\delta\Phi - \delta\rho\partial_i\Phi ,  \label{eq:lin_euler2} \\
    &\partial_i \partial_i \delta\Phi = 4\pi\delta\rho \label{eq:lin_euler3} \,,
\end{align}
where
\begin{align}
    \partial_j\delta\sigma_{ij} ={}& \partial_i(\Gamma_1P\partial_ju_j) + \partial_i(u_j\partial_jP) - \partial_i\left(\frac{2}{3}\mu\partial_ju_j\right) + (\partial_j\mu\partial_j)u_i \notag \\
    &+ \partial_i(u_j\partial_j\mu) - (u_j\partial_j)\partial_i\mu + \mu(\partial_j\partial_ju_i + \partial_i\partial_ju_j) \label{eq:lin_stress} \,.
\end{align}
Note that hereafter the subscript 0 is dropped for unperturbed background quantities.

By combining the perturbed dynamical equations and taking an oscillatory time dependence for all 
perturbed scalar quantities $\delta q (x_i, t) = \delta \bar{q} (x_i) \exp (i \omega t)$ and the
Lagrangian displacement $u_i(x_i,t) = \xi_i(x_i)\exp{(i\omega t)}$, where $\omega$ is the oscillation frequency, one can derive the linear adiabatic wave equation for non-radial oscillations. The angular dependence can be further separated by expanding the perturbed variables in terms of spherical harmonics $Y_{lm}(\theta, \phi)$. In this study, we shall only consider the class of spheroidal modes for which $\xi_i$ and $\delta \bar{q}$ are expanded according to  
\begin{align}
    \xi_r(r,\theta,\phi) &= U(r)Y_{lm}(\theta,\phi) , \label{eq:harmonics_expand_1} \\
    \xi_\theta(r,\theta,\phi) &= V(r)\frac{\partial Y_{lm}(\theta,\phi)}{\partial\theta} \label{eq:harmonics_expand_2} , \\
    \xi_\phi(r,\theta,\phi) &= \frac{V(r)}{\sin{\theta}}\frac{\partial Y_{lm}(\theta,\phi)}{\partial\phi} \label{eq:harmonics_expand_3} , \\
    \delta \bar{q}(r,\theta,\phi) &= \delta \tilde{q}(r)Y_{lm}(\theta,\phi) \label{eq:harmonics_expand_4} . 
\end{align}

Making use of the above expansions, the linear adiabatic wave equation can be recast into a set of differential equations. Let us introduce the variables defined in \citet{Alterman1959} and \citet{Crossley1975}:
\begin{align}
    y_1 &= U , \\
    y_2 &= \lambda X + 2\mu \frac{dU}{dr} , \\
    y_3 &= V , \\
    y_4 &= \mu\left(\frac{dV}{dr}-\frac{V}{r}+\frac{U}{r}\right) , \\
    y_5 &= -\delta\tilde{\Phi} , \\
    y_6 &= -\frac{d\delta\tilde{\Phi}}{dr} - 4\pi\rho U \,,
\end{align}
where $X = \frac{1}{r^2}\frac{d}{dr}(r^2U)-\frac{l(l+1)}{r}V$; $\lambda = \Gamma_1 P - \frac{2}{3}\mu$; $y_1$ and $y_3$ are the radial and transverse displacements, respectively; $y_2$ and $y_4$ are the radial and transverse tractions; $y_5$ and $y_6$ are related to the perturbation of gravitational potential and its derivative.
The differential equations then read \citep{Alterman1959,Crossley1975}
\begin{align}
    y_1' ={}& -\frac{2\lambda\beta}{r}y_1 + \beta y_2 + \frac{l(l+1) \lambda\beta}{r}y_3  ,
    \label{eq:perturb1} \\
    y_2' ={}& \left[-\rho\omega^2-\frac{4\rho g}{r}+\frac{2\delta}{r^2}\right]y_1 - \frac{4\mu\beta}{r}y_2 + l(l+1)\left[\frac{\rho g}{r}-\frac{\delta}{r^2}\right]y_3 \notag \\
    & + \frac{l(l+1)}{r}y_4 - \rho y_6 , \label{eq:perturb2} \\
    y_3' ={}& -\frac{1}{r}y_1 + \frac{1}{r}y_3 + \frac{1}{\mu}y_4 , \label{eq:perturb3} \\
    y_4' ={}& \left[\frac{\rho g}{r}-\frac{\delta}{r^2}\right]y_1 - \frac{\lambda\beta}{r}y_2 + \left[-\rho\omega^2+\frac{\epsilon}{r^2}\right]y_3 - \frac{3}{r}y_4 - \frac{\rho}{r}y_5 , 
     \label{eq:perturb4} \\
    y_5' ={}& 3\gamma y_1 + y_6 , \label{eq:perturb5} \\
    y_6' ={}& -3\gamma\frac{l(l+1)}{r}y_3 + \frac{l(l+1)}{r^2}y_5 - \frac{2}{r}y_6 \label{eq:perturb6} \,,
\end{align}
where $\beta=(\lambda+2\mu)^{-1}$, $\gamma = \frac{4}{3}\pi \rho$, $\delta = 2\mu(3\lambda+2\mu)\beta$, and $\epsilon = 4l(l+1)\mu(\lambda+\mu)\beta - 2\mu$.
Note that in \citet{Crossley1975}, the last term in equation~(\ref{eq:perturb2}) is $-\frac{\rho}{r}y_6$, which is a typo.

In the fluid envelope where $\mu=0$, we have $\beta=\lambda^{-1}$, $\delta=\epsilon=0$, $\lambda=\Gamma_1P$, and $y_4=0$. Therefore, the six differential equations~(\ref{eq:perturb1})-(\ref{eq:perturb6}) reduce to \citep{Alterman1959}
\begin{align}
    y_1' &= -\frac{2}{r}y_1 + \frac{1}{\lambda}y_2 + \frac{l(l+1)}{r}y_3 , \label{eq:fluid1} \\
    y_2' &= -(\rho\omega^2+\frac{4\rho g}{r})y_1 + \frac{l(l+1)\rho g}{r}y_3 - \rho y_6 , \label{eq:fluid2} \\
    y_5' &= 4\pi\rho y_1 + y_6 , \label{eq:fluid3} \\
    y_6' &= -\frac{4\pi\rho l(l+1)}{r}y_3 + \frac{l(l+1)}{r^2}y_5 - \frac{2}{r}y_6 \label{eq:fluid4} \,,
\end{align}
where
\begin{equation}
    \rho\omega^2y_3 = \frac{\rho g}{r}y_1 - \frac{1}{r}y_2 - \frac{\rho}{r} y_5 \label{eq:fluid5} \,.
\end{equation}
This is equivalent to the standard four differential equations \citep[see, e.g.,][]{Cox1980} for a pure fluid star.

To solve the differential equations, one must specify the boundary conditions (BCs) at the center, 
core-envelope interface, and surface. The central BC is obtained by requiring the variables to be regular 
at the center \citep[see][]{Crossley1975}:
\begin{align}
    y_1 &= Ar^{l-1} + A'r^{l+1} , \label{eq:expans1} \\
    y_2 &= Br^{l-2} + B'r^{l} , \label{eq:expans2} \\
    y_3 &= Cr^{l-1} + C'r^{l+1} , \label{eq:expans3} \\
    y_4 &= Dr^{l-2} + D'r^{l} , \label{eq:expans4} \\
    y_5 &= Er^{l} + E'r^{l+2} , \label{eq:expans5} \\
    y_6 &= Fr^{l-1} + F'r^{l+1}  \label{eq:expans6} \,,
\end{align}
where the coefficients $A$, $B$, $C$, etc, are given by
\begin{align}
    A &= lC , \label{eq:expans_coef1} \\
    B &= 2l(l-1)\mu C , \label{eq:expans_coef2} \\
    D &= 2(l-1)\mu C , \label{eq:expans_coef3} \\
    E &= 3\gamma C + \frac{F}{l} \label{eq:expans_coef4} \,,
\end{align}
and
\begin{align}
    C' &= \frac{p_2}{p_1}D' + \frac{\rho}{p_1}\{ F+[\omega^2+(3-l)\gamma]A \} , \label{eq:expans_coef5} \\
    B' &= -q_1 C' + q_2 D' , \label{eq:expans_coef6} \\
    A' &= -lC' + \frac{1}{\mu}D' , \label{eq:expans_coef7} \\
    E' &= \frac{3\gamma}{2(2l+3)}[(l+3)A'-l(l+1)C'] , \label{eq:expans_coef8} \\
    F' &= (l+2)E' - 3\gamma A' \label{eq:expans_coef9} \,,
\end{align}
with
\begin{align}
    p_1 &= 2l^2(l+2)\lambda + 2l(l^2+2l-1)\mu , \\
    p_2 &= l(l+5)+l(l+3)\frac{\lambda}{\mu} , \\
    q_1 &= 2l(l+2)\lambda + 2l(l+1)\mu , \\
    q_2 &= 2(l+1) + (l+3)\frac{\lambda}{\mu} \,.
\end{align}
The above expansions are used in equations~(\ref{eq:perturb1})-(\ref{eq:perturb6}) to start the integration
near the center. 
However, severe numerical cancellation occurs for $y_2'$, $y_4'$, and $y_6'$ when $l \geq 1$. To fix the problem, we need to expand $y_2'$, $y_4'$ and $y_6'$ as well. 
More details about the BCs are presented in Appendix \ref{sec:appendix1}.

At the core-envelope interface, the traction, radial displacement, gravitational potential, and its derivative are continuous, thus one requires all the independent variables (except $y_3$) to be 
continuous \citep{Alterman1959}. The BCs at the stellar surface are $y_2=0$, which can be obtained from the vanishing of the Lagrangian 
perturbation of the pressure, and the continuity of the gravitational potential and its 
derivative
\begin{equation}
    y_6 + \frac{l+1}{r}y_5 = 0 \,.
\end{equation}

\section{Numerical scheme}
\label{sec:numerical_scheme} % used for referring to this section from elsewhere

\subsection{Oscillations}
\label{sec:scheme_osc}

For the central BC, there are 3 independent variables $(C,F,D')$. We choose three orthogonal sets of them, such as $(1,0,0), (0,1,0), (0,0,1)$. After integrating the differential equations, we obtain 3 independent solutions in the solid core, denoted as $\vec{Y_1}(r)$, $\vec{Y_2}(r)$ and $\vec{Y_3}(r)$, where the components of the abstract vectors $\vec{Y_i}$
are $(y_1, y_2, y_3, y_4, y_5, y_6)$. Similarly, we obtain 2 independent solutions $\vec{Y_4}(r)$ and $\vec{Y_5}(r)$ by integrating the differential equations from the surface to the interface in the fluid envelope. The interfacial BC gives the relation between these solutions, namely
\begin{align}
    w_1\vec{Y_1}(R_c) + w_2\vec{Y_2}(R_c) + w_3\vec{Y_3}(R_c) = w_4\vec{Y_4}(R_c) + w_5 \vec{Y_5}(R_c) \,,
\end{align}
for all the components except $y_3$, where the $w$'s are some real constants, and $R_c$ is the radius of the core. This equation is due to the continuity of the $y$'s (except $y_3$). The last coefficient $w_5$ only accounts for an arbitrary normalization and can be set to 1.

The above equation can be rewritten as a matrix equation
\begin{equation}
    \begin{pmatrix}
        Y_{1,1} & Y_{2,1} & Y_{3,1} & Y_{4,1} \\
        Y_{1,2} & Y_{2,2} & Y_{3,2} & Y_{4,2} \\
        Y_{1,4} & Y_{2.4} & Y_{3,4} & Y_{4,4} \\
        Y_{1.5} & Y_{2,5} & Y_{3,5} & Y_{4,5}
    \end{pmatrix}
    \begin{pmatrix}
        w_1 \\
        w_2 \\
        w_3 \\
        -w_4
    \end{pmatrix}
    =
    \begin{pmatrix}
        Y_{5,1} \\
        Y_{5,2} \\
        Y_{5,4} \\
        Y_{5,5}
    \end{pmatrix}
    \label{eq:matching1} \,,
\end{equation}
together with an algebraic equation
\begin{equation}
    w_1Y_{1,6} + w_2Y_{2,6} + w_3Y_{3,6} = w_4Y_{4,6} + Y_{5,6} \label{eq:matching2} \,,
\end{equation}
where $Y_{i,j}$ is the $j$-th component of $Y_i(R_c)$ and we have set $w_5=1$. For a given $\omega$, equation~(\ref{eq:matching1}) determines the coefficients $w_i$, and the value of $\omega$ is a normal mode 
frequency if equation~(\ref{eq:matching2}) is satisfied. We scan through a range of $\omega$ to search for the normal modes.

\subsection{Tidal Love number and deformability}
\label{sec:scheme_tidal}

For the static tide limit we consider in this work, the tidal Love number can be computed using the same set of perturbation equations in Section~\ref{sec:perturbation_eqs} by setting $\omega=0$. This time, the surface boundary condition is replaced by a matching with the exterior solution, thus we integrate from the center up to the surface, in contrast to integrating separately from the center and the surface to the interface, as discussed in Section~\ref{sec:scheme_osc}.

At the center, we again choose 3 orthogonal sets of initial conditions, and obtain 3 independent solutions $\vec{Y}_1(r)$, $\vec{Y}_2(r)$, and $\vec{Y}_3(r)$ by integrating from the center to the core-envelope 
interface. By setting $\omega=0$, we have
\begin{equation}
    y_2 - \rho gy_1 + \rho y_5 = 0
\end{equation}
in the fluid envelope, and $y_4=0$ at the interface. The first equation comes from equation~(\ref{eq:fluid5}), while the second is simply the interfacial boundary condition. These two conditions are enough to determine the solution.

As the general solution is a linear combination of all 3 independent solutions, the above 2 equations can be rewritten as
\begin{align}
    w_1 A_1(R_c) + w_2 A_2(R_c) + w_3 A_3(R_c) &= 0 \label{eq:love_IBC1} \\
    w_1 Y_{1,4}(R_c) + w_2 Y_{2,4}(R_c) + w_3 Y_{3,4}(R_c) &= 0 \label{eq:love_IBC2} \,,
\end{align}
where $A_i(r) = Y_{i,2}(r) - \rho gY_{i,1}(r) + \rho Y_{i,5}(r)$, and the $w$'s are some real constants. The last coefficient $w_3$ is again set to $1$. 
From equations~(\ref{eq:love_IBC1}) and (\ref{eq:love_IBC2}), $w_1$ and $w_2$ are determined, thus the solution at the interface is known, and we can continue integrating towards the surface.

In the fluid envelope with $\omega=0$, it can be shown that equations~(\ref{eq:fluid1})-(\ref{eq:fluid5}) can be combined and reduced to the standard second-order differential equation for determining the tidal Love number in Newtonian gravity~\citep[e.g.,][]{ILoveQ1,Chan2014}
\begin{equation}
\frac{1}{r^2} \frac{d}{dr} \left(r^2 \frac{dH}{dr} \right) - \frac{l(l+1)}{r^2} H
= -4\pi \left( \frac{1}{\rho} \frac{dP}{d\rho} \right)^{-1} H ,
\end{equation}
where $H\equiv \delta {\tilde \Phi}$. 
At the surface, we can define a quantity $\eta = r H'(R)/ H(R)$, where $H(R)$ is obtained by integrating 
the above equation and evaluated at the star surface, and $H'(R)$ is its derivative. 
By matching the interior solution at the surface with the exterior perturbed potential 
(see, e.g., \citet{PoissonWill}), the Love number $k_l$ is determined by 
\begin{equation}
    k_l = \frac{l-\eta}{2(\eta+l+1)} \label{eq:k_l} \,.
\end{equation}
In the discussion of the f-Love relation \citep{Chan2014}, we shall focus on the dimensionless quantity 
$\bar{\lambda}_l \equiv \lambda_l / M^{2l+1}$, where the tidal deformability $\lambda_l$ is related to $k_l$ by
\begin{equation}
\lambda_l = \frac{2}{(2l-1)!!} k_l R^{2l+1} ,
\end{equation}
where $M$ and $R$ are the total mass and radius of the star, respectively.

\section{Results}
\label{sec:results}

In the following, we present our numerical results on oscillations in Section~\ref{sec:results_osc}, tidal deformability in Section~\ref{sec:results_tidal}, and universal relations in Section~\ref{sec:results_relations}. We focus on quadrupolar perturbations so that $l=2$ is assumed for the perturbation equations presented in Section~\ref{sec:perturbation_eqs}. The subscript $l$ in the tidal deformability is thus omitted.

%In Section~\ref{sec:results_osc}, we found that as a star crystallizes from the core, its f-mode encounters a series of avoided crossings, and its f-mode frequency steps up at each avoided crossing. The change in mode frequency is about a few percents, and is larger for stars of lower mass or heavier element. But for i-mode and s-mode, their frequency and frequency spacing decrease with core radius, and are lower for stars of higher mass or lighter element.

%In Section~\ref{sec:results_relations}, the I-Love relation for fluid white dwarf is verified with BPS and our EOS of different elements, but crystallization causes deviation by several percent, thus it could be broken if a white dwarf is crystallized. The f-Love relation \citep{Chan2014} that holds for neutron star is also found to hold for fluid white dwarf, but can be broken when the white dwarf is crystallized.

\subsection{Oscillation modes}
\label{sec:results_osc}

We first discuss the effect of crystallization on the oscillation modes, including the fundamental mode (f-mode), shear mode (s-mode), and interfacial mode (i-mode). The f-mode can be regarded as the fundamental 
pressure mode with no radial nodes. Unlike the f-mode, the s-mode and i-mode do not appear in pure fluid 
stellar models. The former is restored by the elastic stresses in the solid core, and the latter is due to the core-envelope interface. As we assume no composition gradient, there is no gravity mode. Higher-order pressure modes are not considered as well.

The change of pressure mode frequency due to elasticity can be estimated by the $\mu/P$ ratio \citep{Montgomery1999}, which is about a few percent as shown in Figure~\ref{fig:profile}. Therefore, it is expected that the f-mode will also be shifted by a few percent due to crystallization. Moreover, in Figure~\ref{fig:profile}, one can see that the ratio $\mu/P$ is higher at lower density. We found that throughout the density range relevant for WDs, the ratio increases as the density decreases, thus it is expected that the f-mode frequency will show a larger change for stars with lower mass.

\begin{figure}
    \centering
    \includegraphics[width=\columnwidth]{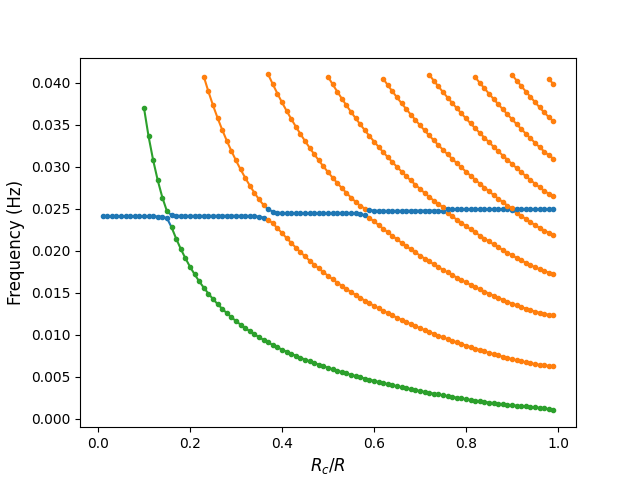}
    \caption{The frequency spectrum of a crystallized $0.20M_\odot$ $^{16}$O star is plotted as a function of 
    the (normalized) core radius. The f-mode, i-mode, and s-modes are denoted by the blue line, green line, and orange lines respectively. Avoided crossing occurs when any two modes are about to "cross" each other.}
    \label{fig:avoided_crossing}
\end{figure}

In Figure~\ref{fig:avoided_crossing}, the frequency spectrum of a $0.20M_\odot$ $^{16}$O star is shown as a function of the solid core radius $R_c$ (normalized by the stellar radius $R$), which is treated as a free parameter in our study. When there is no solid core (i.e., $R_c=0$), the s- and i-modes do not exist, and 
the f-mode is represented by the blue data points in the figure. As the core radius increases, the frequency 
of the f-mode essentially remains unchanged, while the frequency of the i-mode (green data) decreases toward the f-mode. At $R_c/R = 0.15$, there is an avoid crossing between the f-mode and i-mode.  
When $R_c$ increases further, the i-mode keeps decreasing as shown by the lower branch of the green line after the avoid-crossing point. The f-mode remains nearly unchanged (blue line) until another avoid crossing occurs at $R_c/R = 0.37$ between the f-mode and the fundamental s-mode (s$_1$), which is represented by the first branch of orange data points after the green line. 
A few other avoided crossings between the f-mode and higher order s-modes are observed as $R_c/R$ approaches 1. 
Near an avoided crossing, the modes exchange characteristics with each other, and their frequencies deviate from the values they would have if there is no avoided crossing. The deviation is most significant at the crossing between the f and s$_1$ mode. Furthermore, as $R_c/R$ increases from $0$ to $1$, the f-mode frequency increases in steps at each avoided crossing. The step is larger at lower $R_c/R$, but as $R_c/R$ approaches $1$, the step becomes very small. At $R_c/R=1$, however, the i-mode no longer exists, and hence we only show the mode frequencies up to $R_c/R=0.99$.

\begin{figure}
    \centering
    \includegraphics[width=\columnwidth]{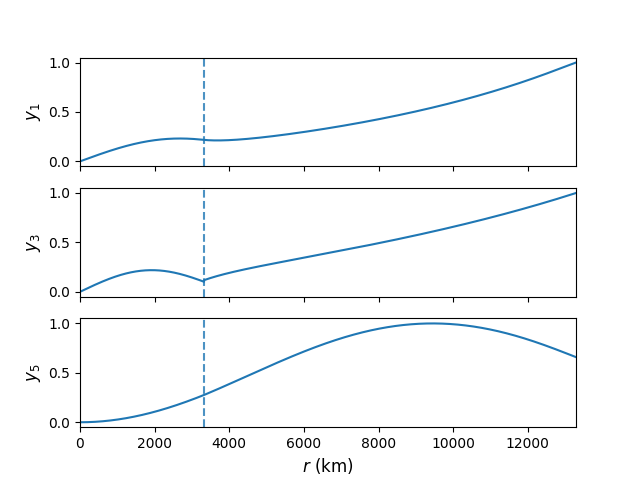}
    \caption{Eigenfunctions for the f-mode of a $0.20M_\odot$ $^{16}$O star with crystallized core radius $R_c=0.25R$, normalized by their maximum amplitudes. The core-envelope interface is denoted by the dashed line.}
    \label{fig:eigenfunc_f}
\end{figure}

\begin{figure}
    \centering
    \includegraphics[width=\columnwidth]{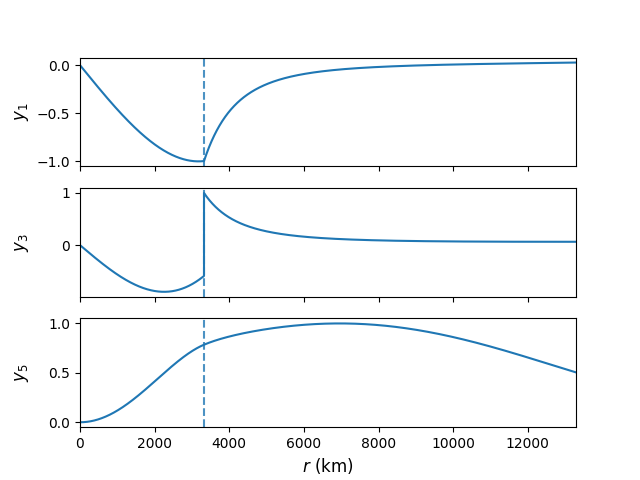}
    \caption{Similar to Figure~\ref{fig:eigenfunc_f}, but for the i-mode.}
    \label{fig:eigenfunc_i}
\end{figure}

\begin{figure}
    \centering
    \includegraphics[width=\columnwidth]{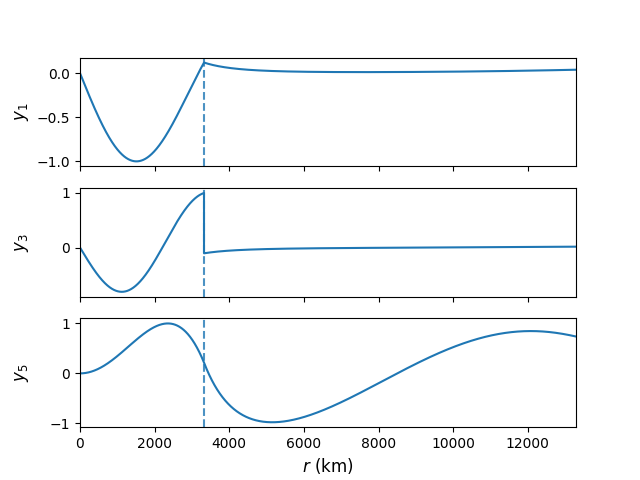}
    \caption{Similar to Figure~\ref{fig:eigenfunc_f}, but for the s$_1$-mode. }
    \label{fig:eigenfunc_s1}
\end{figure}

Next, we show the eigenfunctions $y_1$, $y_3$, and $y_5$ for different oscillation modes.
Let us recall that $y_1$ and $y_3$ are the radial and transverse displacements, respectively; $y_5$ is related to the potential perturbation. 
In Figures~\ref{fig:eigenfunc_f}, \ref{fig:eigenfunc_i}, and \ref{fig:eigenfunc_s1}, the eigenfunctions of the f-mode, i-mode, and s$_1$-mode are presented respectively. All of them are computed for a $0.20M_\odot$ $^{16}$O WD model with $R_c/R=0.25$, and normalized by their own maximum amplitudes. In each figure, the vertical dashed line represents the position of the core-envelope
interface. The characteristic of f-mode is the overall increasing trend of the fluid displacements ($y_1$ and $y_3$). In particular, the fluid displacements increase 
monotonically toward the star surface in the fluid envelope just like the behavior for 
a pure fluid star. Inside the solid core, the displacements may be oscillating, but the overall increasing trend remains. The discontinuity in the transverse displacement ($y_3$) is also small for the f-mode. 
On the other hand, the i-mode is characterized by the large displacements at the interface, as well as the large discontinuity in the transverse displacement. The potential perturbation $y_5$ for the i-mode is continuous across the core-envelope interface as it should be. 

For the s$_1$-mode in Figure~\ref{fig:eigenfunc_s1}, the main feature is the strong oscillations in the solid core. The order of the s-mode in general can be told by the number of nodes in the displacements in the solid core. For instance, the s$_1$ mode has one node in both $y_1$ and $y_3$ in the core. However, the displacements
are small in the fluid envelope. Its potential perturbation may continue oscillating in the fluid layer, and the oscillation depends on the frequency of the s-mode.

The general characteristics of the different modes discussed above remain the same when $R_c/R$ increases. 
The eigenfunctions presented in \Cref{fig:eigenfunc_f,fig:eigenfunc_i,fig:eigenfunc_s1}
are somewhat different from those in \citet{Fuller2014}, where a fluid giant planet with a solid core is considered. They found that the f-mode is confined to the fluid layer while the s-modes are confined to the solid core. However, we show that the f-mode can have nonzero oscillations in the solid core, while the s$_1$-mode oscillations can extend 
somewhat outside the core. The difference may be due to the fact that there is a density jump at the 
core-envelope interface in the giant planet model which does not exist in our models\footnote{As a verification of our numerical code, we have also calculated the 
oscillation modes of the giant planet model in \citet{Fuller2014} and found that our results agree with those reported by them (see Appendix~\ref{sec:appendix2}). }.

\begin{figure}
    \centering
    \includegraphics[width=\columnwidth]{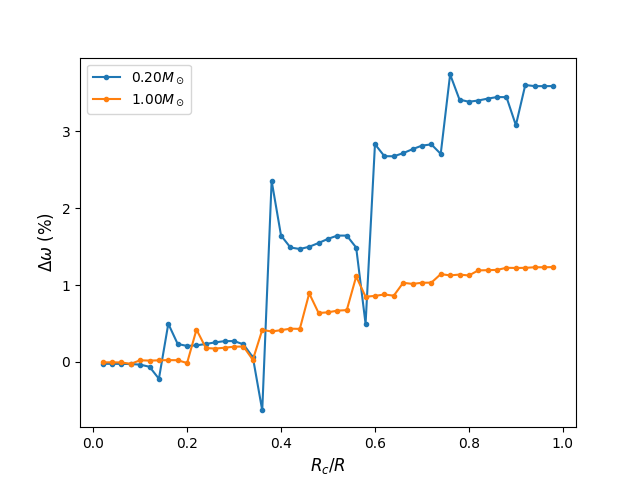}
    \caption{Percentage change of the f-mode frequency $\Delta\omega = (\omega-\omega_\text{fluid})/\omega_\text{fluid}$ due to the presence of a solid core 
    is plotted against the (normalized) solid core radius $R_c/R$ for two $^{16}$O stars of different masses. The kinks are due to avoided crossings between the f-mode and other modes as illustrated in Figure~\ref{fig:avoided_crossing}.}
    \label{fig:f_varying_mass}
\end{figure}

\begin{figure}
    \centering
    \includegraphics[width=\columnwidth]{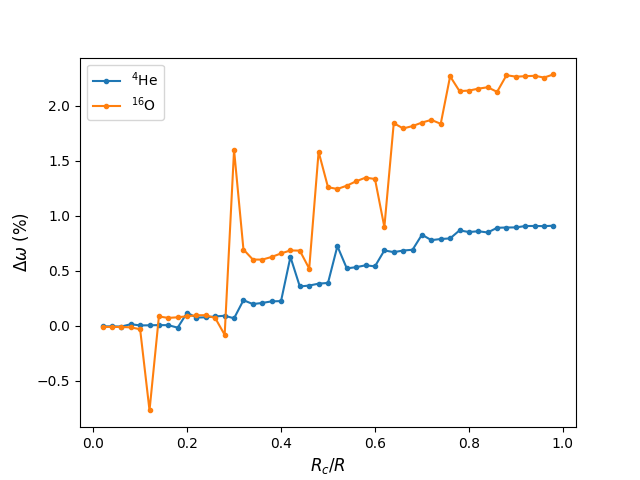}
    \caption{Similar to Figure~\ref{fig:f_varying_mass}, but for two $0.40M_\odot$ stars with different elements. The kinks are due to avoided crossings between the f-mode and other modes.}
    \label{fig:f_varying_element}
\end{figure}

After examining the oscillation modes qualitatively, we investigate quantitatively how 
the mode frequencies change as the size of the crystallized core increases. Figure~\ref{fig:f_varying_mass} plots the percentage change of the f-mode frequency $\Delta\omega = (\omega-\omega_\text{fluid})/\omega_\text{fluid}$ against
the (normalized) core radius $R_c/R$ for two $^{16}$O stars of masses $0.20 M_\odot$ and $1.00 M_\odot$, where $\omega$ is the frequency for a crystallized WD model and $\omega_\text{fluid}$ is the frequency for the corresponding 
pure fluid counterpart. The spikes in the curves are due to avoided crossings between the f-mode and other modes
(see Figure~\ref{fig:avoided_crossing}). 
The f-mode frequency generally increases with the solid core radius, and jumps in step at each avoided crossing, similar to Figure~\ref{fig:avoided_crossing}. In agreement with the estimation based on the $\mu/P$ ratio, the mode frequency indeed changes more significantly for the less massive star. For heavier stars, avoided crossing happens more frequently as the crystallized core grows, but the step sizes and spikes are smaller. 

Similarly, Figure~\ref{fig:f_varying_element} compares the percentage change of the f-mode frequency for two $0.40M_\odot$ stars made of $^4$He and $^{16}$O respectively. Larger frequency changes are observed for stars of heavier elements as expected, since $\mu$ is proportional to $Z^{2/3}$. On the other hand, more frequent avoided crossings with smaller step sizes and spikes are found for stars of lighter elements. In general, if a white dwarf is partially or fully crystallized, the f-mode frequency may deviate from its pure fluid value by a few percent.

\begin{figure}
    \centering
    \includegraphics[width=\columnwidth]{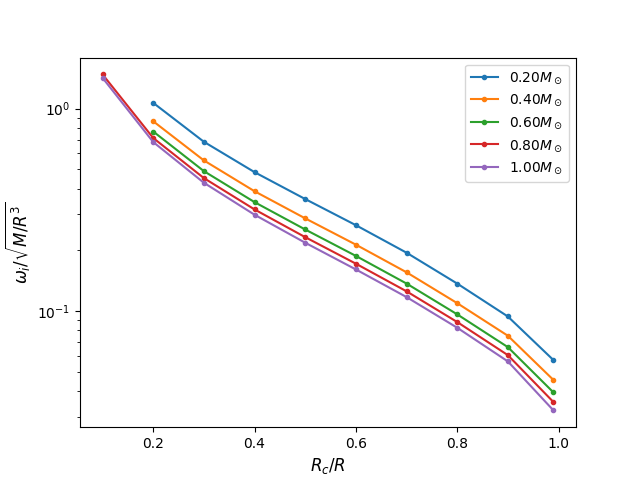}
    \caption{Comparison of the i-mode frequencies (normalized by $\sqrt{M/R^3}$) for $^{16}$O stars of different masses.}
    \label{fig:i_varying_mass}
\end{figure}

\begin{figure}
    \centering
    \includegraphics[width=\columnwidth]{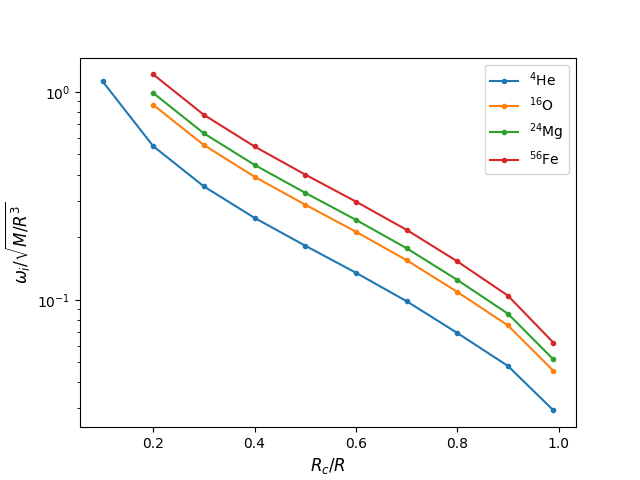}
    \caption{Similar to Figure~\ref{fig:i_varying_mass}, but for $0.40M_\odot$ stars of different elements.}
    \label{fig:i_varying_element}
\end{figure}

In contrast, the i-mode frequency shows a simple decreasing trend as the solid core radius increases. Figure~\ref{fig:i_varying_mass} plots the i-mode frequencies for $^{16}$O stars of different masses against 
$R_c/R$. For better visualization of the data, we normalize the i-mode frequencies by the characteristic 
frequency $\sqrt{M/R^3}$ in the figure as the models have large differences in their masses and radii.
It is seen that the i-mode frequency decreases as the core radius increases along a fixed mass sequence.  
The normalized i-mode frequency also decreases as the mass increases for a fixed value of $R_c/R$.
Note, however, that the original (unnormalized) i-mode frequency actually increases with the mass. 
For instance, the i-mode frequency increases from 4.49 mHz to 23.7 mHz as the mass increases from $0.20 M_\odot$
to $1.00 M_\odot$ for $R_c/R =0.6$.
Four $0.40M_\odot$ stars of different elements are also compared in Figure~\ref{fig:i_varying_element}, showing that the normalized i-mode frequency for lighter element is lower as well for a fixed value of $R_c/R$. In this case, 
the trends of the original and normalized i-mode frequencies are the same. 
Again, at $R_c/R=1$, the i-mode does not exist and we only show the mode frequencies up to $R_c/R=0.99$. If $R_c/R$ increases further towards 1, the i-mode frequency drops rapidly to 0.
Overall, the i-mode frequency can decrease to a few percent of $\sqrt{M/R^3}$ for $R_c/R \leq 0.99$.
%However, if there is a density jump at the core-envelope interface, the behavior of the i-mode frequency could be different.

\begin{figure}
    \centering
    \includegraphics[width=\columnwidth]{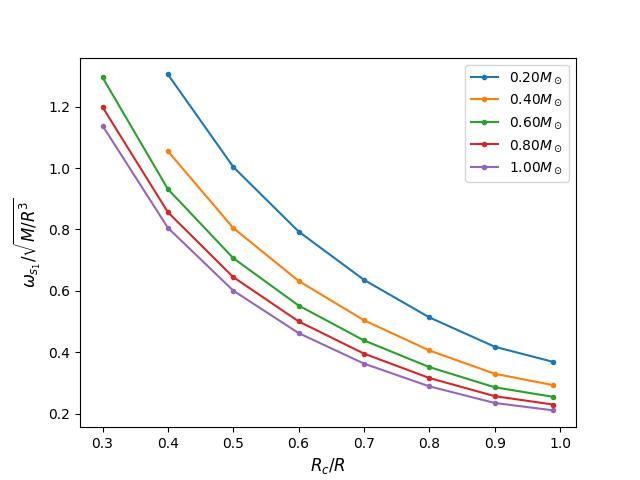}
    \caption{Comparison of the s$_1$-mode frequencies (normalized by $\sqrt{M/R^3}$) for $^{16}$O stars of different masses.}
    \label{fig:s_varying_mass}
\end{figure}

\begin{figure}
    \centering
    \includegraphics[width=\columnwidth]{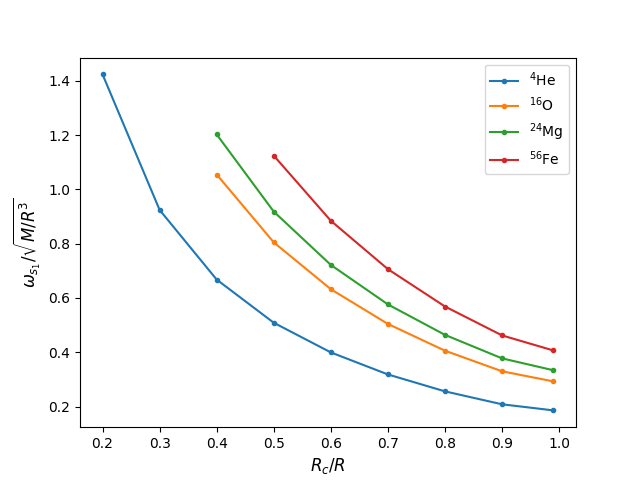}
    \caption{Similar to Figure~\ref{fig:s_varying_mass}, but for $0.40M_\odot$ stars of different elements.}
    \label{fig:s_varying_element}
\end{figure}

For the s-mode, we show the frequencies of the s$_1$-mode for $^{16}$O stars of different masses and for $0.40M_\odot$ stars of different elements in Figures~\ref{fig:s_varying_mass} and \ref{fig:s_varying_element}, respectively. The general trend is similar to that of the i-mode. The mode frequency decreases as the core radius increases for a fixed mass sequence. The mode frequency is also decreased for lighter elements.

In summary, the f-mode frequency increases with the solid core radius in steps due to the phenomenon of  
avoided mode crossings. The steps are larger for stars that are less massive or made of heavier elements, but are more frequent for stars that are massive or made of lighter elements. For the i-mode and s-mode, their frequencies decrease monotonically with the solid core radius.

\subsection{Tidal deformability}
\label{sec:results_tidal}

\begin{figure}
    \centering
    \includegraphics[width=\columnwidth]{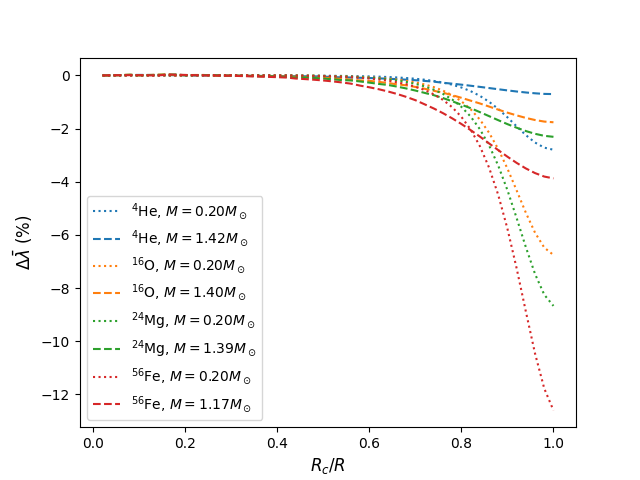}
    \caption{Percentage change of the scaled tidal deformability $\Delta\bar{\lambda}=(\bar{\lambda}-\bar{\lambda}_\text{fluid})/\bar{\lambda}_\text{fluid}$ as a function of the solid core radius, for different models near the maximum or minimum mass.}
    \label{fig:love_deviation}
\end{figure}

After studying the oscillation modes, we now turn our attention to the effect of crystallization on the tidal deformability. In Figure~\ref{fig:love_deviation}, we show how the scaled tidal deformability 
$\bar{\lambda}=\lambda/M^5$ deviates from its value in the pure fluid limit as the core radius increases. 
In the figure, the percentage change $\Delta \bar{\lambda} = (\bar{\lambda}-\bar{\lambda}_\text{fluid})/\bar{\lambda}_\text{fluid}$ is plotted against $R_c/R$ for WD models with different
masses and elements. For instance, the blue dotted line represents the results for a $^4$He star with 
$0.20 M_\odot$. It is seen that $\Delta \bar{\lambda}$ is very small when $R_c < 0.5 R$ for all models. 
Its magnitude increases to only about 1\% level even when the core radius increases to $R_c \approx 0.7R$. 
Hence, the fluid envelope can screen out the effect of the solid core significantly. 
This screening effect is present regardless of the element and mass of the models, though it is slightly weaker when the star is more massive. 
Such a screening effect due to the fluid envelope has also been seen in the study of the tidal deformability for hybrid star models featuring a crystalline color-superconducting quark-matter core \citep{LauSY:2019}. 
The change in $\bar{\lambda}$ becomes significant and depends sensitively on the star models when $R_c \geq 0.7 R$. For a given element, the deviation increases as the mass decreases. For a $0.20 M_\odot$ $^{56}$Fe star, the magnitude 
of $\Delta \bar{\lambda}$ can increase to more than 10\% for a fully crystallized WD.

\begin{figure}
    \centering
    \includegraphics[width=\columnwidth]{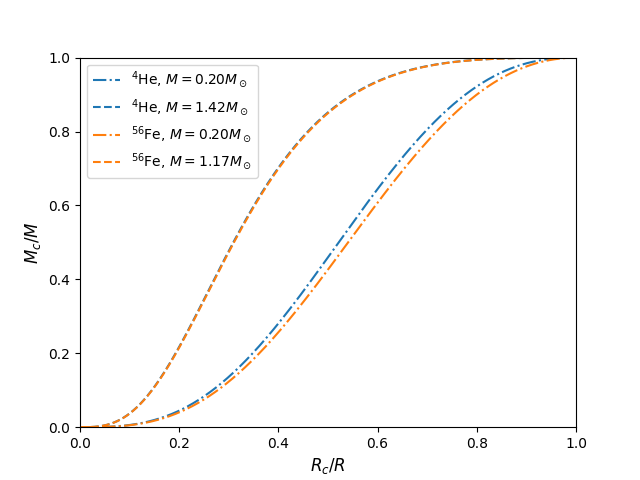}
    \caption{Relations between the crystallized mass fraction $M_c/M$ and the normalized core radius $R_c/R$ for
     $^4$He and $^{56}$Fe stars with different masses.}
    \label{fig:mr}
\end{figure}

While we characterize the extent of crystallization by the (normalized) solid core radius $R_c/R$, observational studies often relate the effect by crystallized mass fraction $M_c/M$ instead. To demonstrate the relation 
between the two quantities, we show $M_c/M$ as a function of $R_c/R$ for models of the heaviest ($^{56}$Fe) and lightest ($^4$He) elements in Figure~\ref{fig:mr}. 
The curves for the two low-mass ($0.20 M_\odot$) stars are close to each other. We also observe that more massive 
stars achieve higher $M_c/M$ at lower $R_c/R$ for both elements. In particular, the two high-mass cases $1.42M_\odot$ ($^4$He) and $1.17M_\odot$ ($^{56}$Fe) shown in the figure are stars near the Chandrasekhar limit of the corresponding elements. It is interesting to see that the two curves agree very well in this limit. We notice that the corresponding curves for all the WD models considered in this study bascially lie in the region between the two orange lines in the figure.

\subsection{Universal relations}
\label{sec:results_relations}

\begin{figure}
    \centering
    \includegraphics[width=\columnwidth]{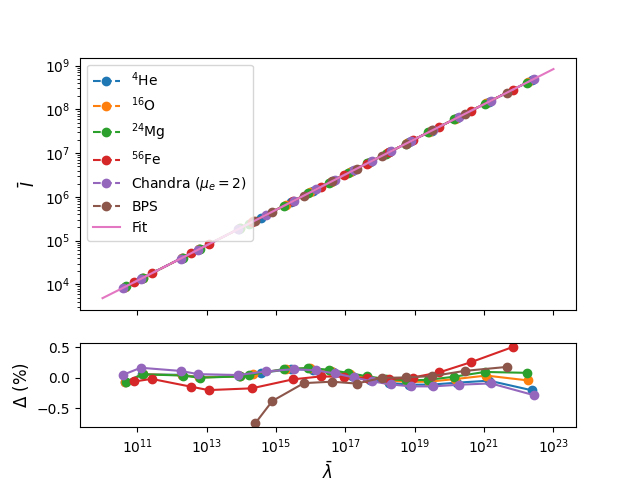}
    \caption{The I-Love relation for pure fluid WDs without solid cores. The numerical data are obtained from our chosen EOS models
    (see text for details). The solid line is the fitting curve given by equation~(\ref{eq:ILove_fit}).  
    The percentage difference between the data and the fitting curve is shown in the lower panel.
    Note that the sequence for the BPS EOS is truncated close to the maximum-mass stable WD model 
    where $\bar\lambda \approx 10^{14}$.    }
    \label{fig:Ilove1}
\end{figure}

Here we shall study the effects of crystallization on the I-Love and f-Love universal relations for WDs. While 
the I-Love-Q relations for pure fluid WD models have been studied in \citep{Boshkayev2017,Boshkayev2018,Taylor2020}, 
we shall first use our own EOS models to verify the I-Love relation in the fluid limit. 
The moment of inertia $I$ is simply calculated from the equilibrium fluid background, and hence is unaffected by the solid core. In Figure~\ref{fig:Ilove1}, we plot the scaled moment of inertia ${\bar I}\equiv I/M^3$ against 
$\bar \lambda$ for pure fluid WD models using our chosen EOS models discussed in Section~\ref{sec:model}, which include the Chandrasekhar model with the mean molecular weight per electron $\mu_e = 2$ \citep{Chandrasekhar1931,Chandrasekhar1939}, the BPS EOS \citep{BPS1971}, and the model described by 
equations~(\ref{eq:eos1})-(\ref{eq:eos2}) with different elements. 
We fit our numerical results using a 4th order polynomial fitting curve (solid line)
of $\ln{\bar{I}}$ versus $\ln{\bar{\lambda}}$ which is given by
\begin{align}
    \ln{\bar{I}} &= 0.710 + 0.233\ln{\bar{\lambda}} + 0.00691(\ln{\bar{\lambda}})^2 \notag \\
    &\quad - 1.20\times 10^{-4}(\ln{\bar{\lambda}})^3 + 7.51\times 10^{-7}(\ln{\bar{\lambda}})^4 .
    \label{eq:ILove_fit}
\end{align}
The percentage deviation of the numerical data from the fitting curve is shown in the lower panel. It is seen that
the I-Love relation is insensitive to the EOS models to within 1\% level. The BPS model shows a larger deviation 
toward smaller values of $\bar{\lambda}$. Toward this end, the BPS model is reaching the maximum-mass instability limit. 
We note that the I-Love relation for pure fluid WDs has been provided before by \citet{Boshkayev2018}, but their data is truncated to 2 significant figures only. 
As the effect of crystallization on the I-Love relation is at most a few percent level (see below), 
we thus present a more accurate fitting curve for the I-Love relation in the fluid limit for the following 
investigation.

\begin{figure}
    \centering
    \includegraphics[width=\columnwidth]{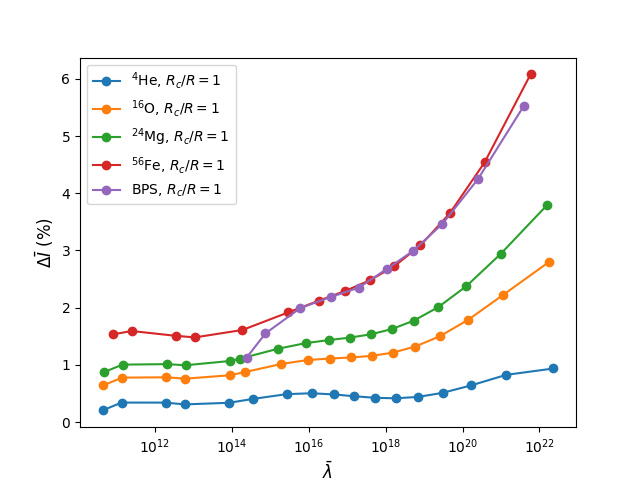}
    \caption{The percentage deviations of the scaled moment of inertia for crystallized WDs with solid core radius $R_c/R=1$ from the prediction of the I-Love relation (equation~(\ref{eq:ILove_fit})) for pure fluid WDs. }
    \label{fig:Ilove2}
\end{figure}

We now use the I-Love relation in the fluid limit (i.e., equation~(\ref{eq:ILove_fit}))
as a benchmark to compare with the results obtained by crystallized WD models. We define the 
percentage deviation $\Delta {\bar I} \equiv ({\bar I} - {\bar I}_\text{Fit})/ {\bar I}_\text{Fit}$, 
where $\bar I$ is the moment of inertia of a crystallized WD and ${\bar I}_\text{Fit}$ is the 
corresponding value predicted by equation~(\ref{eq:ILove_fit}) at the same ${\bar \lambda}$. 
In Figure~\ref{fig:Ilove2}, we show $\Delta {\bar I}$ for fully crystallized WD models.
It should be recalled that the moment of inertia is calculated from the fluid background star and is unchanged by crystallization. It is in fact the change in the tidal deformability due to the solid core 
for a given background model that produces the deviation.
Note also that $\bar \lambda$ decreases with increasing stellar mass, and hence the region below
$\bar \lambda \approx 10^{16}$ is occupied by WDs near the maximum mass limits. For these 
lower values of $\bar \lambda$, the percentage deviation $\Delta {\bar I}$ is below 2 percent. 
For higher values of $\bar \lambda$, corresponding to lower mass WDs, the deviation can be as
large as about 5\% for some models. 

As a result, crystallization can break the I-Love relation of WDs, and the deviation is more significant 
for low mass WDs, though fully crystallized stars 
are needed to reach a deviation up to a few percent level.

\begin{figure}
    \centering
    \includegraphics[width=\columnwidth]{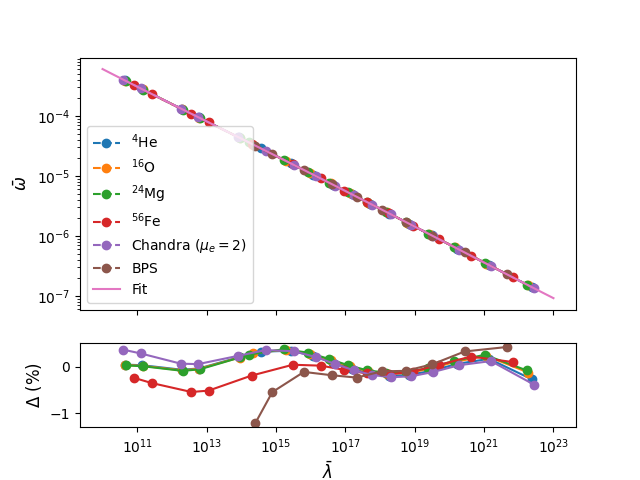}
    \caption{The f-Love relation for pure fluid WDs. The solid line is the fitting curve given
    by equation~(\ref{eq:fLove_fit}). The percentage difference between the data and the fitting curve is shown in the lower panel.}
    \label{fig:flove1}
\end{figure}

After studying the effects of crystallization on the I-Love relation, we now turn to the f-Love relation which 
was originally discovered for neutron stars \citep{Chan2014} and connects the scaled f-mode frequency ${\bar \omega} \equiv \omega M$ to $\bar \lambda$. 
As far as we are aware, the corresponding f-Love relation for pure fluid WDs has not been studied before 
and we shall fill this gap in the literature by establishing such a relation for WDs. 
In Figure~\ref{fig:flove1}, we plot $\bar \omega$ against $\bar \lambda$ for pure fluid WDs using our chosen EOS models. It is seen clearly that the results in fact lie on an approximate EOS-insensitive universal relation 
just like in the case of neutron stars. The solid line in the figure is a fitting curve given by 
\begin{align}
    \ln{\bar{\omega}} &= 1.66 - 0.570\ln{\bar{\lambda}} + 0.0119(\ln{\bar{\lambda}})^2 \notag \\
    &\quad - 2.16\times 10^{-4}(\ln{\bar{\lambda}})^3 + 1.40\times10^{-6}(\ln{\bar{\lambda}})^4 .
    \label{eq:fLove_fit}
\end{align}
The lower panel of Figure~\ref{fig:flove1} shows the percentage deviation of the numerical data from the 
fitting curve and it is seen that the accuracy is within 1\% level. Again, the BPS EOS shows
a larger deviation when the stellar models approach the maximum-mass instability limit
at smaller value of $\bar{\lambda}$.

\begin{figure}
    \centering
    \includegraphics[width=\columnwidth]{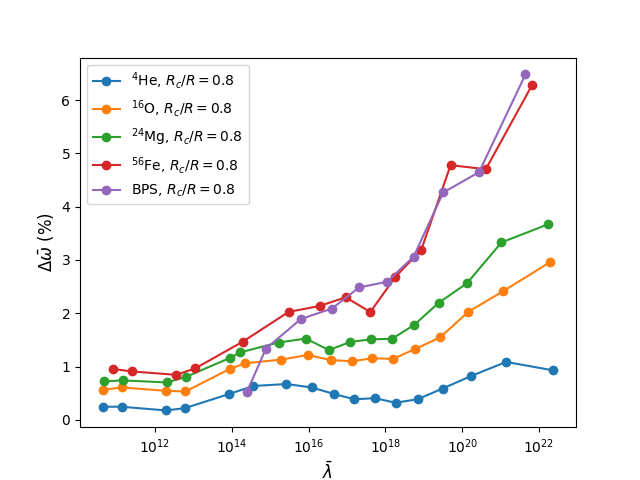}
    \caption{Percentage deviations from the f-Love relation (i.e., equation~(\ref{eq:fLove_fit})) for stars with solid core radius $R_c/R=0.8$. Similar to \Cref{fig:f_varying_mass,fig:f_varying_element}, the kinks are due to the phenomenon of avoided mode crossings.}
    \label{fig:flove2}
\end{figure}

However, similar to the I-Love relation, the f-Love relation is broken for crystallized WDs. 
As in the study of the breaking of the I-Love relation due to crystallization, we here define the percentage
deviation $\Delta {\bar \omega} \equiv ({\bar \omega} - {\bar\omega}_\text{Fit}) / {\bar \omega}_\text{Fit}$,
where $\bar\omega$ is the scaled f-mode frequency of a crystallized WD and ${\bar\omega}_\text{Fit}$ is the corresponding value predicted by equation~(\ref{eq:fLove_fit}) at the same $\bar\lambda$. 
In Figure~\ref{fig:flove2}, we plot $\Delta\bar{\omega}$ for crystallized WDs with solid core radius $R_c/R=0.8$ against $\bar\lambda$ for our EOS models. 
As in the case of the I-Love relation, the effects of crystallization is small (within about 1\% level) for 
smaller values of $\bar\lambda$, corresponding to the region near the maximum mass limits. 
The deviations become more significant as $\bar\lambda$ increases in the low-mass region. For instance, 
the percentage deviation for the BPS model with $0.2 M_\odot$ reaches up to 6.5\%. 
It is seen that there are kinks at various points along each sequence in Figure~\ref{fig:flove2}. They 
are due to the phenomenon of avoided mode crossings as discussed above. 
Similar to the breaking of I-Love relation, crystallization causes deviation from the f-Love relation by a few percent. The deviation is larger for star models composed of heavier elements or in the low-mass region with
larger $\bar\lambda$. 

\begin{figure}
    \centering
    \includegraphics[width=\columnwidth]{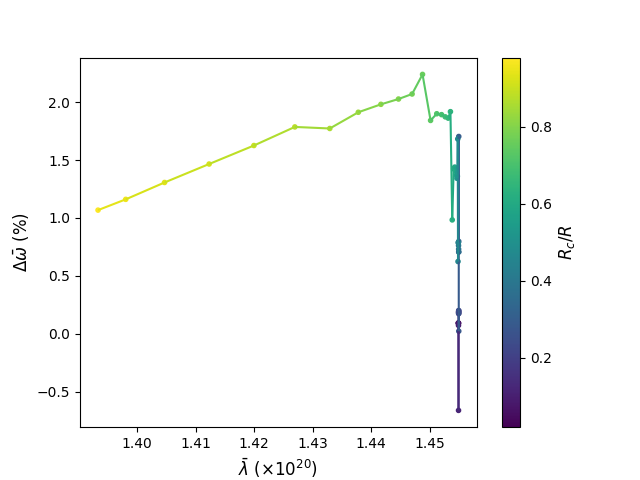}
    \caption{The trajectory of a $0.40M_\odot$ $^{16}$O star in the $\Delta\bar{\omega}-\bar{\lambda}$ space. As the solid core radius $R_c/R$ increases, its f-mode frequency increases and deviates from the f-Love relation. At higher $R_c/R$, $\bar{\lambda}$ quickly decreases, reducing the deviation from the f-Love relation. The kinks on the trajectory are due to the phenomenon of avoided mode crossings, and $\Delta\bar{\omega}$ does not start from $0$ since the f-Love relation is only a fitting curve. } 
    \label{fig:flove3}
\end{figure}

If the core size grows beyond $0.8R$, one might guess that the deviations would further increase. However, we found that this is not the case. In Figure~\ref{fig:flove3}, we demonstrate how a $0.40M_\odot$ $^{16}$O star moves in the $\Delta\bar{\omega}-\bar{\lambda}$ plane as the core radius $R_c/R$ grows from $0$ to $1$. 
It should be noted that
the trajectory of the star does not start with $\Delta\bar{\omega}=0$ when $R_c=0$ since the f-Love relation is fit to the numerical data of different stellar models to about 1\% only. 
When $R_c/R$ is smaller than $0.5$, the f-mode frequency increases and deviates from the f-Love relation, while $\bar{\lambda}$ remains unchanged due to the screening effect discussed in Section~\ref{sec:results_tidal}. The kinks at various points are due to the phenomenon of avoided mode crossings. But when $R_c/R$ becomes larger, the deviation from the f-Love relation decreases. This is because $\bar{\lambda}$ starts decreasing rapidly as shown in Figure~\ref{fig:love_deviation}, moving the value closer to the f-Love relation.

\section{Conclusions}
\label{sec:conclusions}

The GW signals emitted from binary WD systems at the merger phase may be detectable by future space-borne GW detectors \citep{Kinugawa2022}. During the late inspiral when the WDs are close to each other, the tidal 
deformation and excitation of the f-mode oscillations of the WDs become important and contribute to the 
GW signals. The knowledge of the oscillation modes and tidal deformability of WDs is important to 
interpreting the observed GW signals. 
On the other hand, it is predicted that WDs in binaries may have evolved long enough to be fully or partially crystallized \citep{vanHorn1968} and strong observational evidence has been found recently \citep{Tremblay2019}. It is thus important to study how crystallization inside WDs may affect the f-mode
frequency and tidal deformability, and hence the emitted GW signals.

%Determining the signal waveform requires knowledge about the oscillation modes of the WDs. Universal relations will also play an important role in inferring physical properties of these WDs. While both the oscillation modes and universal relations may be altered by crystallization, there remains limited investigations into their effect. We have therefore analyzed the effect of crystallization on the oscillation modes of WDs and their universal relations.

In this paper, we have studied not only the f-mode of WDs, but also the s- and i-modes which appear only when a crystalline core is present. For a given background WD model, we find that the frequencies of the s- and
i-modes decrease as the radius of the core increases, while the f-mode frequency is affected only slightly.
However, various avoided mode crossings occur when the frequencies of the f-mode and other modes
approach each other as the core size increases. The f-mode frequency generally increases with the core 
radius and jumps in steps at the avoided crossing points. We also find that crystallization has a larger 
effect on the f-mode frequency for low mass WDs. For instance, the percentage change of the f-mode frequency
for a fully crystallized $^{16}$O WD (comparing to its pure fluid counterpart) increases from about 1\% to 
3\% when the mass of the star decreases from $1.00 M_\odot$ to $0.20 M_\odot$. 

The tidal deformability of crystallized WDs is also studied in this work. Due to a screening effect of 
the fluid envelope, the effect of the crystalline core in general becomes significant only when the core radius $R_c \geq 0.7 R$. Similar to the case of f-mode, the effect increases as the mass decreases. For 
instance, the tidal deformability of a $0.2 M_\odot$ $^{56}$Fe fully crystallized WD can be different from 
that of a pure fluid counterpart by about 12\%. 

Similar to the situation for neutron stars, there exist approximate EOS-insensitive universal relations connecting the moment of inertia, tidal deformability, and f-mode frequency for pure fluid WDs. 
We have used our EOS models to verify the I-Love relation proposed by \citet{Boshkayev2017} and present 
a more accurate fitting curve for the study. We have also established the f-Love relation for fluid 
WDs for the first time. 
By taking the effects of elasticity in the calculations of the f-mode frequency and tidal 
deformability, we find that the universal relations can be broken for crystallized WDs. 
While the deviations are only about 1\% level in the high-mass region near the maximum mass limits, 
the deviations for lower mass WDs can be around 5 percent for some models. 
Therefore, crystallization may affect the results when universal relations are employed to infer the properties of WDs,
such as the individual masses of 
the system, from the GW signals emitted from a WD binary \citep{Wolz2021}. 
On the other hand, the universal relations may be used to reduce the matter variables in theoretical waveform modellings for binary WDs. We may thus (in principle) infer the properties of crystallized WDs by comparing observation data against waveform models with and without the assumption of universal relations, though very 
accurate GW measurements would be needed. 

However, as pointed out by \citet{Perot2022}, fully ionized matter with completely degenerate electron gas at zero temperature is assumed in the EOS presented in Sec.~\ref{sec:model}. This approximation only holds in the crystallized core of WDs. Yet, for WDs that are cold enough to crystallize, we expect the deviation to be small in the fluid envelope as well.
On the other hand, general relativistic effects are important for WDs near the maximum mass limit \citep{Perot2022},
thus the universal relations established by us in Newtonian gravity may be modified for high-mass WDs near this limit. These issues are left for a future investigation.

\section*{Acknowledgements}
This work is partially supported by a grant from the Research Grant Council of the Hong Kong Special Administrative
Region, China (Project No. 14300320). 

\section*{Data availability}
The data underlying this article are available in the article. No new data needed to be generated or analyzed.

%%%%%%%%%%%%%%%%%%%%%%%%%%%%%%%%%%%%%%%%%%%%%%%%%%
%\section*{Data Availability}
%The inclusion of a Data Availability Statement is a requirement for articles published in MNRAS. Data Availability Statements provide a standardised format for readers to understand the availability of data underlying the research results described in the article. The statement may refer to original data generated in the course of the study or to third-party data analysed in the article. The statement should describe and provide means of access, where possible, by linking to the data or providing the required accession numbers for the relevant databases or DOIs.



\bibliographystyle{mnras}
\bibliography{example} % if your bibtex file is called example.bib

%\begin{thebibliography}{99}
%\end{thebibliography}

%%%%%%%%%%%%%%%%%%%%%%%%%%%%%%%%%%%%%%%%%%%%%%%%%%

%%%%%%%%%%%%%%%%% APPENDICES %%%%%%%%%%%%%%%%%%%%%

\appendix

\section{Boundary conditions at the center}
\label{sec:appendix1}

Near the center, the $y$'s can be expanded using coefficients presented in \citet{Crossley1975}, and the derivatives can be obtained by plugging the expansions into \cref{eq:perturb1,eq:perturb2,eq:perturb3,eq:perturb4,eq:perturb5,eq:perturb6}. However, severe numerical cancellation occurs among $y_2'$, $y_4'$, and $y_6'$ when $l \geq 1$. To fix this issue, we evaluate the expansion of their derivatives near the center according to 
\begin{align}
    y_2' ={}& 2\mu l(l-1)(l-2)Cr^{l-3} + (-\rho\omega^2-4\rho\gamma)Ar^{l-1} \notag \\
    & + l(l+1)\rho\gamma Cr^{l-1} - \rho Fr^{l-1} + 2\delta A'r^{l-1} - 4\mu\beta B'r^{l-1} \notag \\
    & - l(l+1)\delta C'r^{l-1} + l(l+1)D'r^{l-1} \\
    y_4' ={}& 2\mu(l-2)(l-1)Cr^{l-3} + lD'r^{l-1} + \rho\gamma A'r^{l+1} - \rho\omega^2 C'r^{l+1} \notag \\
    & - \rho E'r^{l+1} \\
    y_6' ={}& (l-1)Fr^{l-2} - 3\gamma l(l+1) C'r^l + l(l+1)E'r^l - 2F'r^l \,,
\end{align}
where the coefficients $A$, $B$, $C$, etc, are given by \cref{eq:expans_coef1,eq:expans_coef2,eq:expans_coef3,eq:expans_coef4,eq:expans_coef5,eq:expans_coef6,eq:expans_coef7,eq:expans_coef8,eq:expans_coef9}. For $y_1'$, $y_3'$, and $y_5'$, \cref{eq:perturb1,eq:perturb3,eq:perturb5} can be applied directly.

When $r$ is small, we integrate the $y$'s using the above expansions. But when $r$ is larger such that the numerical cancellation is under control, we switch back to \cref{eq:perturb1,eq:perturb2,eq:perturb3,eq:perturb4,eq:perturb5,eq:perturb6}.

\section{Code tests}
\label{sec:appendix2}

The calculation of oscillation modes is divided into the crystalline core and fluid envelope. We have checked that our code can reproduce the f-mode frequency of 
a $n=1$ polytropic star (i.e. $P \propto \rho^2$) in the pure fluid limit \citep[see Table 17.2b of][]{Cox1980} to within $0.1\%$ accuracy when either the core radius is set to be very small (e.g., $R/R_c =0.01$) or the shear modulus of a finite-size core is set to be much smaller than the central pressure.

%%%%%%%%%%%%%%%%%%%%%%%%
%\begin{table}
%    \centering
%    \caption{The f-mode frequency of a polytropic star $P=k\rho^2$ with $R_c/R=0.01$ (near fluid limit), compared to the purely fluid result in \citet{Cox1980}. For the treatment of elasticity, we considered 3 cases, $\mu=0.1P$, $\mu=0.05P$ and $\mu=0.01P$.}
%    \begin{tabular}{ccc}
%        \hline
%        $\mu/P$ & $\omega^2R^3/M$ (our work) & $\omega^2R^3/M$ (Cox)\\
%        \hline
%        0.1 & 1.4977 & 1.498 \\
%        0.05 & 1.4977 & 1.498 \\
%        0.01 & 1.4977 & 1.498 \\
%        \hline
%    \end{tabular}
%    \label{tab:cox_test}
%\end{table}
%%%%%%%%%%%%%%%%

To further validate our oscillation code, we test it using the 
Saturn model in \citet{Fuller2014}, which is constructed from a $n=1$ polytropic model by multiplying a factor of $4$ to the density in the core ($r<0.25R$), and normalizing 
the density profile such that the mass and radius equal to those of Saturn. In contrast to our white dwarf models, there is a density jump at the 
core-envelope interface in their Saturn model which we have also taken into account in the test. 
Since \citet{Fuller2014} did not report a precise value of the radius of their model, we chose a model that best matches their density, bulk modulus and gravity profiles. We have checked that our code accurately reproduces the f-mode, s$_1$-mode, and i-mode frequencies as presented in Table~\ref{tab:jupiter_test}. Note that the numerical results of 
\citet{Fuller2014} are extracted from their figures.

\begin{table}
    \centering
    \caption{Comparison between f-mode, s$_1$-mode and i-mode frequency of the Saturn model in \citet{Fuller2014} computed by our code and computed by \citet{Fuller2014}.}
    \begin{tabular}{ccc}
        \hline
        Mode & $\sqrt{\omega^2R^3/M}$ (Our work) & $\sqrt{\omega^2R^3/M}$ (Fuller)\\
        \hline
        f & 1.316 & 1.312\\
        s$_1$ & 0.411 & 0.413\\
        i & 2.773 & 2.770\\
        \hline
    \end{tabular}
    \label{tab:jupiter_test}
\end{table}

%%%%%%%%%%%%%%%%%%%%%%%%%%%%%%%%%%%%%%%%%%%%%%%%%%

% Don't change these lines
\bsp	% typesetting comment
\label{lastpage}
\end{document}